\documentclass[10pt, dvipsnames]{wlscirep}
\usepackage[utf8]{inputenc}
\usepackage[T1]{fontenc}

\title{Automated Contact Tracing: a game of big numbers in the time of COVID-19}

\author[1,2,3,*,$\dag$]{Hyunju Kim}
\author[4,5,*,$\dag$]{Ayan Paul}

\affil[1]{Beyond Center for Fundamental Concepts in Science, Arizona State University, Tempe, AZ, USA}
\affil[2]{ASU-SFI Center for Biosocial Complex Systems, Arizona State University and Santa Fe Institute, USA}
\affil[3]{School of Earth and Space Exploration, Arizona State University, Tempe, AZ, USA}
\affil[4]{DESY, Notkestra{\ss}e 85, D-22607 Hamburg, Germany}
\affil[5]{Institut f\"ur Physik, Humboldt-Universit\"at zu Berlin, D-12489 Berlin, Germany}
\affil[*]{e-mail: \href{mailto:hyunju.kim@asu.edu}{hyunju.kim@asu.edu}, \href{mailto:ayan.paul@desy.de}{ayan.paul@desy.de}}
\affil[$\dag$]{These authors contributed equally to this work.}

\keywords{COVID-19, SARS-CoV-2, Contact Tracing, Disease Mitigation}

\begin{abstract}
One of the more widely advocated solutions for slowing down the spread of COVID-19 has been automated contact tracing. Since proximity data can be collected by personal mobile devices, the natural proposal has been to use this for automated contact tracing providing a major gain over a manual implementation. In this work, we study the characteristics of voluntary and automated contact tracing and its effectiveness for mapping the spread of a pandemic due to the spread of SARS-CoV-2. We highlight the infrastructure and social structures required for automated contact tracing to work. We display the vulnerabilities of the strategy to inadequate sampling of the population, which results in the inability to sufficiently determine significant contact with infected individuals. Of crucial importance will be the participation of a significant fraction of the population for which we derive a minimum threshold.  We conclude that relying largely on automated contact tracing without population-wide participation to contain the spread of the SARS-CoV-2 pandemic can be counterproductive and allow the pandemic to spread unchecked. The simultaneous implementation of various mitigation methods along with automated contact tracing is necessary for reaching an optimal solution to contain the pandemic.
\end{abstract}

\begin{document}
\flushbottom
\maketitle

\thispagestyle{empty}
\rhead{DESY 20-069/HU-EP-20/10}

\section*{Introduction}

A relentless and damaging battle is being fought against the spread of COVID-19. While several countries have managed to significantly slow down its spread, severe measures have had to be taken to do so and at great cost to the economic and social well-being of the nations. It is still not certain when a significant control over the spread of SARS-CoV-2 can be attained. Recent projections propose surveillance for the next few years~\cite{Kisslereabb5793}, with several measures that will need to be put in place to minimize the cost of the pandemic to humankind. Automated contact tracing is one of these measures.

Contact tracing has been observed to be effective in previous pandemics (or epidemics) like the Ebola virus outbreak in 2014-2015~\cite{10.1371/journal.pntd.0006762}. This preemptive method allows for the containment of the pathogen by isolating potentially infected individuals that have been traced. Extensive studies of manual contact tracing were done during the previous outbreak of the Ebola virus~\cite{doi:10.1142/S1793524518500936,BROWNE201533,1551-0018_2018_5_1165}, SARS-CoV and MERS-CoV~\cite{KWOK2019186}. More recently, mathematical models have been formulated to study contact tracing assuming the disease spread to be quantifiable by the SIR model~\cite{OKOLIE2020108320}. However, the efficacy of automated contact tracing during the SARS-CoV-2 pandemic requires a more detailed examination given the distinct difference in the prevalence of this pandemic from the ones in the recent past and the different modes of transmission of the pathogen. 


Manual contact tracing is not very effective against pathogens that spread like the influenza virus but is more effective for containing smallpox and SARS-CoV and partially effective in containing foot-and-mouth disease~\cite{10.1371/journal.pone.0000012}. The viral shedding patterns of SARS-CoV and MERS-CoV are similar~\cite{10.1093/cid/civ951,Chowell:2015tn} and show almost no presymptomatic transmission~\cite{Fraser6146},\footnote{One study suggested that MERS-CoV can be transmitted before the onset of symptoms~\cite{Omrani:2013fp}.} while Ebola is known to be transmitted through the bodily fluids of infected individuals after the onset of symptoms~\cite{REWAR2014444}. On the other hand, influenza shows a significant rate of viral shedding in the presymptomatic stage~\cite{Lau:2010tv}. The important transmission characteristics of SARS-CoV-2 that set it apart from other HCoV pathogens like SARS-CoV and MERS-CoV and from Ebola are:
\begin{itemize}
\setlength\itemsep{0em} 
    \item SARS-CoV-2 transmission is driven by presymptomatic spreading like the influenza virus~\cite{Santarpia2020.03.23.20039446,Wang:2020jb,He:2020ty}. 
    \item The pathogen can be transmitted through the air in high contamination regions and through contaminated dry surfaces for several days~\cite{Santarpia2020.03.23.20039446,doi:10.1056/NEJMc2004973,Guo:2020ww} leading to its high transmission rates. This brings about additional challenges when the disease cannot be contained within an isolated envelope of a healthcare system. While a similar spreading pattern is seen in SARS-CoV and MERS-CoV, this makes SARS-CoV-2 more easily transmittable than Ebola.
    \item The ACE2 binding of SARS-CoV-2 is estimated to be relatively stronger than SARS-CoV and might explain its observed spreading characteristics~\cite{Zhou:2020wk,Wrapp1260,CHEN2020135}.
\end{itemize}
In the early stages of the pandemic the reproduction number $R_0$, for SARS-CoV-2 was estimated to be $2.2 - 2.7$~\cite{Lieabb3221,doi:10.1056/NEJMoa2001316,Wu:2020ix,10.2807/1560-7917.ES.2020.25.4.2000058,Du:2020va}, similar to SARS-CoV.\footnote{Much higher reproductive rates have also been estimated with data from Wuhan, China~\cite{Sanche:2020wm}. In general there are variabilities in the estimation of the reproduction number with time and containment strategies~\cite{ALRAEEI2020}.} The dispersion parameter is estimated to also be similar to that of SARS-CoV (close to 0.1), which could be causing superspreading~\cite{LloydSmith:2005ue,10.2807/1560-7917.ES.2020.25.4.2000058,Hellewell:2020ek,10.12688/wellcomeopenres.15842.1}. 

In principle, automated contact tracing can be shown as an effective means of containing SARS-CoV-2~\cite{Hellewell:2020ek}. However, factors such as long delays from symptom onset to isolation, fewer cases ascertained by contact tracing, and increasing presymptomatic transmission can significantly impact how effective automated contact tracing will be in practice. Normally a significant contact is defined as  being within 2 meters for at least 15 minutes~\footnote{cf. \href{https://www.cdc.gov/coronavirus/2019-ncov/php/contact-tracing/contact-tracing-plan/contact-tracing.html}{https://www.cdc.gov/coronavirus/2019-ncov/php/contact-tracing/contact-tracing-plan/contact-tracing.html}}. Keeling et al. demonstrated that this can result in the detection of more than 4 out of 5 secondary infections but at the cost of tracing 36 contacts per individual~\cite{Keeling2020.02.14.20023036}.  Changes to the definitions of a significant contact can reduce the numbers traced. For example, if the minimum time required to be considered a significant contact is increased, the number of people needed to be traced will decrease at the cost of not being able to identify potentially infected individuals. Detailed modeling of SARS-CoV-2 transmission shows that the pandemic can be sustained just by presymptomatic transmission and that automated contact tracing can be used to contain the spread of the pathogen if there are no significant delays to identifying and isolating infected individuals and their contacts~\cite{Ferrettieabb6936}. 

Considering all the factors that make contact tracing a different game for SARS-CoV-2, in this paper we will examine in detail how much data and participation from the population will be needed to make automated contact tracing effective. This will give an estimate of the necessary scale of implementation of automated contact tracing and whether it will be feasible. The model that we build with parameters that are mostly independent of each other or factorized, will also allow for the estimation of the effects of various mitigation methods like the use of personal protective equipment (PPE) in enhancing the efficacy of automated contact tracing which we discuss before the discussion section. In this work we address voluntary and automated contact tracing using proximity data alone excluding methods such as the use of CCTV, credit card information, logging of identities of individuals during vists to locations and travels, etc. that have been successfully used by many countries like Singapore~\cite{Pung:2020ef}, Taiwan~\cite{info:doi/10.2196/19540}, South Korea~\cite{10.1001/jama.2020.6602} and China~\cite{Ferrettieabb6936} for contact tracing.

\section*{Contact tracing for COVID-19}

To judge the efficiency of contact tracing, it is crucial to determine whether an infectious disease can spread in the presymptomatic stage or from asymptomatic individuals. For a disease that can spread only in the symptomatic stage, the infected individuals can spread the disease to their contacts before they are isolated and to medical workers after they are isolated with varying probabilities. Of significance here is that after the initial period of ignorance of the population about a rising pandemic, infected individuals will be isolated with higher efficiency (even with manual contact tracing) resulting in the curtailment of the spread of the pathogen. How is contact tracing more effective in such diseases? Since the mobility of the infected individual usually sees a decline after the onset of symptoms, the number of contacts at risk become limited to only those who are most often in contact with the individuals and hence traceable manually. 
This allows the implementation of a manual contact tracing algorithm that identifies these neighbors and isolates or tests them as suggested in reference ~\cite{10.1371/journal.pone.0000012}. This was seen to be effective during the Ebola, MERS-CoV and SARS-CoV outbreaks.

However, the spreading of SARS-CoV-2 follows a very different pattern. With the prevalence of spreading of infection through presymptomatic and subclinical hosts, the number of individuals that might need to be traced can be very large. This has led to the belief that automated contact tracing in a wider gamut should be implemented. Most of the proposed solutions~\cite{Hellewell:2020ek,Keeling2020.02.14.20023036,Ferrettieabb6936} require the use of historical proximity data to trace contacts. In the context of COVID-19, there are some obvious pitfalls in the algorithm:

\begin{itemize}
\setlength\itemsep{0em}
    \item It is estimated that about 86\% (95\% CI: [82\% -- 90\%]) of the infected cases in China were undocumented prior to the travel ban on the 23$^{rd}$ of January 2020 generating 79\% of the documented infections~\cite{Lieabb3221}. A large number of these undocumented cases experienced mild, limited or no symptoms and can hence go unrecognized. Similar results were reported by other studies~\cite{Daym1375,Verity:2020cu}. It is not possible to trace all the contacts of these individuals since they will be partially reported leading to incomplete coverage of contact tracing. 
    \item While it is assumed that the SARS-CoV-2 spreads within a proximity radius of $r_0$ (assumed to be 2 meters), not much is known about the probability of transmission, $p_t$ when two individuals come within this domain of contact for a minimum contact time $t_0$. Assuming $p_t$ to be large will lead to an unreasonably large estimate of the number of potential infections required to be traced in a crowded region like supermarkets, which remain open even during the period of social distancing. On the other hand, assuming $p_t$ to be small will underestimate the number of infected contacts, especially because there might be other modes of transmission of SARS-CoV-2 that are not being considered. By definition $p_t$ depends on the dynamics of disease transmission when a healthy individual comes in significant contact with a infected individual. Moreover, $p_t$ is not constant over $r_0$ and also varies with the stage of infection the infected individual is at~\cite{He:2020ty}. Several other factors contribute to the value of $p_t$ in addition to the contagiousness of the disease including, but not limited to, the use of PPE, public awareness of the disease, whether the surrounding is open or a closed area, air circulation (freely circulating as opposed to air conditioned), etc.~\cite{MORAWSKA2020105832}.  
\end{itemize}

The first pitfall can be alleviated by increasing the testing rate of individuals for viral RNA in the hope that a larger fraction of the asymptomatic or mildly symptomatic carriers can be traced. Increasing awareness can also help. The second pitfall can be alleviated when more detailed knowledge of the spread of SARS-CoV-2 is available and with the help of simulation of the spread of the disease in a population. For the rest of the work, we will assume $p_t$ to be a variable and $r_0$ to be fixed to 2 meters~\cite{Keeling2020.02.14.20023036}. 

The real-world applicability of automated contact tracing requires the examination of the effects of partial sampling of the population. The assumption that we are working with is that enrollment in automated contact tracing will be voluntary and individuals remain free to do one of the following:
\begin{itemize}
\setlength\itemsep{0em}
    \item Choose not to enroll in the program by either not using the application or the devices needed for tracing, including discontinuity in participation.
    \item Choose not to report on their health condition which is assumed to be voluntary.
\end{itemize}
Both types of occurrences have an effect of reducing the efficacy of automated contact tracing but in slightly different manners. In the first case, not subscribing to the service would not only remove an individual from the pool that is being notified, but it also removes them from the pool of individuals that are reporting. In the second case only the latter happens.

\section*{Modeling automated contact tracing}

\begin{figure}
    \centering
    \includegraphics[width=0.95\textwidth]{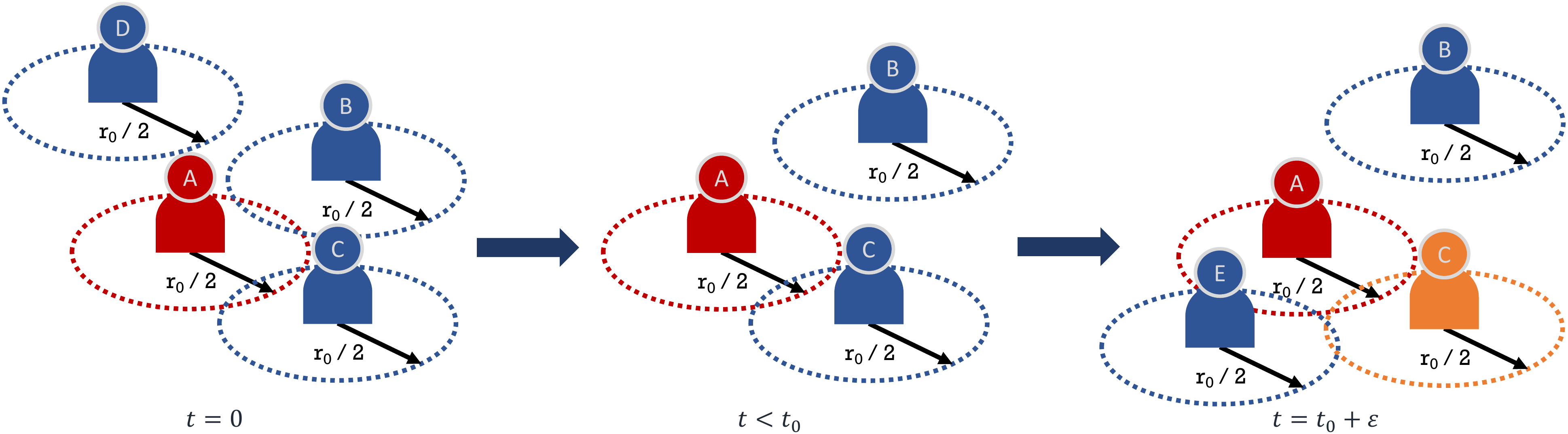}
    \caption{A depiction of automated contact tracing. The cross-section is denoted by the dashed circle and is of radius $r_0/2$. Interactions occur from $t=0$ to $t=t_0+\epsilon$ where $\epsilon \ll t_0$. $A$ will be confirmed as COVID-19 positive in the future and $C$ will be notified having come in contact with $A$. $E$ might be notified if $E$ stays in contact with $A$ for a time period greater than $t_0$.}
    \label{fig:interaction}
\end{figure}

Since, in automated contact tracing, a significant contact has to be less than $r_0$ distance away for time $t_0$, we describe every individual by a circle with a radius of $r_0/2$ which we shall call the cross-section of the individual. The cross-section is chosen such that any overlap between two cross-sections can be taken as a significant contact between the two respective individuals. Temporally, the cross-sections have to overlap for a time $t_0$ which is the threshold interaction time that is assumed critical for an individual to infect another by proximity. 
For the sake of simplicity and with some loss of generality of our argument, we can assume that the probability of getting infected, $p_t$, is independent of the degree of overlap of the cross-sections\footnote{This assumption is to mimic how automated contact tracing is implemented through mobile devices where the probability of infection is not considered as variation depending on distance between two users as long as it is closer than a physical proximity threshold set by the contact tracing. In addition, $p_t$ as a function of distance is not very well known as yet.} and for any time $t>t_0$ as is done normally in automated contact tracing. 

Figure~\ref{fig:interaction} gives a depiction of what automated contact tracing would be for a group of individuals. 
In the left-most panel, $B$ and $C$ are in contact with $A$ at $t=0$ but not with each other. $D$ is isolated from all of them. After a period of time $t<t_0$, $B$ is isolated but $C$ stays in contact with $A$. Then at time $t=t_0+\epsilon$, where $\epsilon\ll t_0$, we see that $C$ is still in contact with $A$, $B$ remains isolated and $E$ has come in contact with both $A$ and $C$. Using the methods of automated contact tracing, if $A$ reports as being tested as infected withing 14 days of the encounter with $C$, $C$ will be deemed as having had significant contact with $A$. $E$ might also be deemed as such depending on how long he maintains proximity with $A$, but the proximity of $E$ with $C$ need not be counted even if $E$ spends $t>t_0$ in contact with $C$ (if only primary contacts are traced) unless $C$ reports as being infected too. 

This method of automated contact tracing will work as long as $A$ and $C$ (and possibly $E$) are enrolled in the service even if $B$ and $D$ are not. However, $D$ is completely isolated and by remaining so for a long time is observing social distancing from any other individual. $B$ is representative of an individual who observes partial social distancing. Hence, for $D$ this service is not necessary and for $B$ it is of limited value. If $C$ is not enrolled in the service $C$ will never get notified if $A$ gets tested as infected. $C$ might fall get confirmed as infected or become an asymptomatic carrier and continue contaminating others. If $A$ does not enroll in this service then $C$ never gets notified leading to the same conclusions but $E$ might get notified if $C$ declares being infected and $E$ is enrolled in the service. 

An estimated 45\% of person-person virus transmissions occur from individuals who are in the presymptomatic phase~\cite{Ferrettieabb6936}. Prevalence of subclinical infections of SARS-CoV-2 further reduces the effectiveness of contact tracing. 
With automated contact tracing using a definition of $r_0 = 2$ meters and $t_0 = 15$ minutes more than 80\% of the cases can be traced~\cite{Keeling2020.02.14.20023036} if every infected case is reported. 
In what follows we create a simplified model of automated contact tracing to deduce the minimum fraction of the population that needs to enroll in the program for it to be effective.
\begin{itemize}
\setlength\itemsep{0em}
    \item Let $N$ be the number of individuals in a population and $f_i$ the fraction of the population that is infected, regardless of whether they know it or not. Therefore, the true number of infected individuals is $f_i N$.
    \item If testing is conducted only when mild or severe symptoms are seen (i.e. excluding testing of asymptomatic cases), the number of confirmed cases is $r_cf_i N$ with $r_c$ being the fraction of the infected that will be confirmed as infected by testing.
    \item We define $f_e$ as the fraction of the population that is enrolled for automated contact tracing and $f_c$ as the fraction of the users that will confirm that they have been diagnosed positive. Hence, the number of individuals that have tested positive, are using automated contact tracing and will confirm that they have been tested as infected is $f_cf_er_cf_iN$. 
    \item We define $a_c$ as the average number of individuals that a contagious individual has significant contact with over the period in which they are contagious, significant contact being defined as lasting for a period of time greater than $t_0$ and within a radius of $r_0$. The period over which an individual is contagious is about 5 days on an average for those who spread the disease in the presymptomatic phase and can be longer for asymptomatic and sub-clinical cases~\cite{Ferretti2020.09.04.20188516}.
 \end{itemize}
 Since only $f_e$ fraction of contacts are using the service, we can estimate the number of individuals that can be traced as $f_cf_er_cf_iNa_cf_e$. Note that we assumed $f_e$ and $f_c$ are uniform even though  $r_cf_iN$ is not a random sample of the overall population with the purpose of estimating the number in the most conservative scenario. In the real world, $f_e$ will be likely lower amongst the set of individuals that actually get infected (and their immediate contacts) and higher in the conjugate set due to different levels of caution exercised by the two groups, which, in turn, results in the decrease of the number of traceable contacts.\footnote{The variations in $f_e$ within different demographic groups are not accounted for in our work and this can potentially be correlated with the way the disease spreads. Here $f_e$ is the fraction of the whole population that continuously use the service. Accounting for these variations within our model is possible but requires more data.}
 
To compute the number of individuals that need to be quarantined or isolated since they are now at risk of being infected from coming in contact with a contagious person, we define the following.
 \begin{itemize}
 \setlength\itemsep{0em}
    \item Since $p_t$ is defined as the probability of transmission of infection within the proximity radius $r_0$ being exposed for a time greater than $t_0$, the number of individuals who are potentially newly infected is, on average, $p_t f_i N a_c $, i.e., $p_t$ multiplied by the number of contacts of the group of infected individuals.\footnote{Here we make a simplifying assumption that the disease has spread to only a small fraction of the population and the probability of a single healthy individual to randomly have significant contact with two contagious individuals in a period of 14 days is negligibly small in general. There will be outliers depending on the habits of individuals but we can neglect them for this analysis.}
    \item Finally, we define $f_T$ as the fraction of the individuals at risk of being infected that needs to be successfully quarantined to quell the spread of the pathogen. In addition to other factors, $f_T$ also depends on the delay in isolating potentially infected individuals~\cite{Ferrettieabb6936}.
 \end{itemize} 
Therefore, the number of individuals that should be quarantined is $f_T p_t f_iNa_c$. For automated contact tracing to work effectively, we have,
\begin{equation}
f_e^2f_cr_cf_iNa_c \geq f_T p_t f_iNa_c.  
\label{eq:need_can}
\end{equation}

\section*{The game of big numbers}

\begin{figure}[t]
    \centering
    \includegraphics[width=0.8\textwidth]{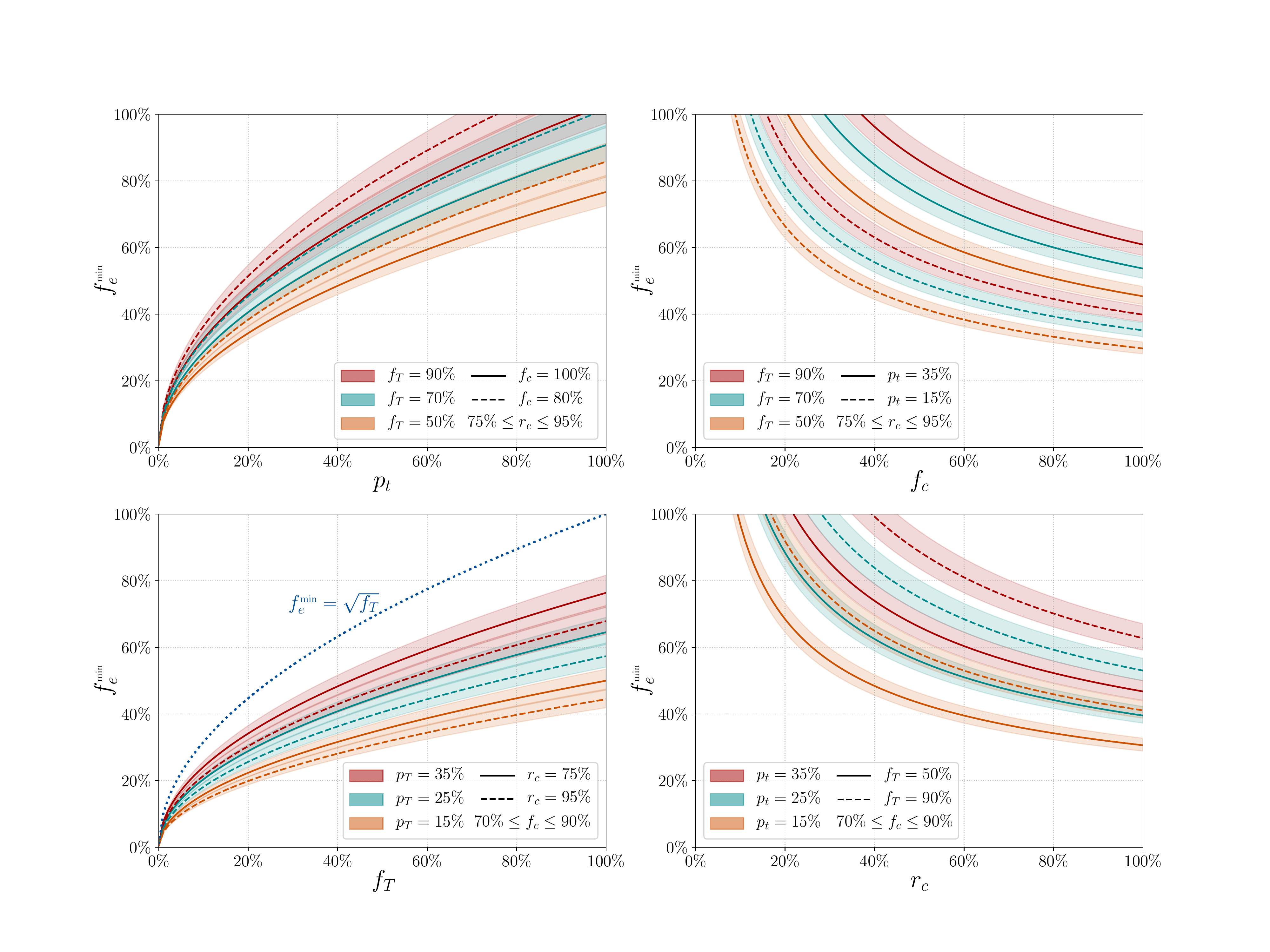}
    \caption{Percentage of the population that needs to be enrolled ($f^{min}_e$) for automated contact tracing to be successful. Starting from the left, the solid and dashed lines represent $f_c = 100\%, 80\%$ respectively for the first panel, $p_t = 35\%, 15\%$ for the second panel,  $r_c = 75\%, 95\%$ for the third panel and $f_T = 50\%, 90\%$ for the fourth panel. For the left two panels the fraction of truly infected individuals that will be confirmed as infected by testing, $r_c$ is varied between $75\% - 95\%$. For the right two panels the fraction of people who will confirm they have been tested as infected if they are enrolled, $f_c$ is varied between $70\% - 90\%$. Three cases for the minimum fraction of the individuals at risk that need to be traced are considered with $f_T=50\%,70\%,90\%$ in orange, green and red respectively in the left two panels and similarly, three cases are considered for $p_t=15\%,25\%,35\%$ in the right two panels. The blue dotted line in the third panel from the left gives the threshold variation of $f^{min}_e$ with $f_T$ when all other parameters are set to 1. The y axes are identical for all panels. See text for more details.}
    \label{fig:f_e_p_t}
\end{figure}

Eq.~(\ref{eq:need_can}) simply states that the number of individuals that can be notified by automated contact tracing (on the left-hand side) has to be greater than or equal to the number of individuals who need to be notified (on the right-hand side). Note that $a_c$, the average number of contacts, drops out of the inequality and hence, the inequality is independent of the population density of the region since eq.~(\ref{eq:need_can}) is in terms of fraction of the population and not the absolute number of individuals. This simply implies that in a region of denser population a larger number of people need to be contacted and quarantined but leaves $f_e$ independent of the population density. Since the right-hand side is the minimum fraction of the population that needs to be traced we arrive at:
\begin{equation}
f^{min}_e = \sqrt{\frac{f_Tp_t}{f_cr_c}}.
\label{eq:min_enroll}
\end{equation}
The fraction $f^{min}_e$ is the minimum fraction of the population that needs to be enrolled in automated contact tracing for it to be effective as a means of slowing down the spread of the pandemic. In eq.~(\ref{eq:min_enroll}), $p_t$ depends on the spreading dynamics of the pathogen determined by individual-to-individual interactions and, therefore, also depends on the mitigating measures taken at both the population level and the individual level. Naively, in automated contact tracing, $p_t$ is taken as one if the contact has lasted for over time $t_0$ with the subjects being less than $r_0$ apart. This can be reduced by use of PPE or other mitigation methods as we discuss later. The parameter $f_T$ depends on the disease spreading dynamics and can be estimated from modeling the disease spreading amongst a population~\cite{Ferrettieabb6936}. From both Hellewell et. al.~\cite{Hellewell:2020ek} and Ferretti et. al.~\cite{Ferrettieabb6936} it is seen that 60\% - 80\% of the contacts need to be successfully traced and quarantined instantly to contain the outbreak over a period of time which makes $f_T\sim 0.6 - 0.8$. The slower the response to the identification of contact at risk higher is $f_T$ for the same reduction rate of the reproduction number. We assume that identification of contact at risk takes less than a day in automated contact tracing. The parameter $r_c$ is governed by the ability to identify infected individuals through testing and depends on the protocols of the testing program and its coverage. On the other hand, $f_c$ is determined solely by the degree to which individuals are willing and able to confirm that they have been tested positive.

Let us examine the limit $p_t=f_c=r_c=1$. This is the limit where every significant contact is assumed to be at risk, everyone who is enrolled in the automated contact tracing program reports as infected when tested positive and every infected individual can be successfully identified by testing. Then we arrive at the relation $f^{min}_e = \sqrt{f_T}$ (blue dotted line in the third from left panel of figure~\ref{fig:f_e_p_t}). Since $f_T$ is the fraction of contacts that need to be successfully isolated, it can be extracted from the abscissa of fig. 3 of ref.~\cite{Ferrettieabb6936}. For example, if 100\% of the infected cases can be isolated, then for a change in the epidemic growth rate by -0.1, one needs $f_T\sim 60\%$. Hence $f^{min}_e \sim 77\%$. It is intuitive that $f^{min}_e$ scales as the squareroot of $f_T$ since both the infected and the contact at risk need to be enrolled and the probability that each are enrolled is $f_e$ leading to $f_T\propto f_e^2$. It gives the threshold which $f^{min}_e$ cannot exceed for any given $f_T$.

Lastly, we define the effectiveness of the automated contact tracing, $\eta$, as the ratio of the actual number of individuals that will be notified ($f_e^2f_cr_cf_iNa_c$) to the minimum number of individuals that should be notified to quell the spread of the disease (${f^{min}_e}^2f_cr_cf_iNa_c$) and get:
\begin{equation}
\eta \equiv \frac{f_e^2}{{f^{min}_e}^2}.    
\end{equation}

Figure~\ref{fig:f_e_p_t} depicts how the fraction of the population that needs to be enrolled for the automated contact tracing program to be successful ($f^{min}_e$) varies with the four factorized parameters. In the left-most panel of figure~\ref{fig:f_e_p_t}, we show the minimum percentage of the population that needs to be enrolled in automated contact tracing $f^{min}_e$ (in \%) versus the transmission probability $p_t$. 
We consider two values for $f_c = 80\%$, 100\%, the fraction of individuals who test positive and will confirm their symptoms to trigger automated contact tracing, by the solid and dashed lines respectively. The solid and dashed lines represent $p_t=15\%,25\%$ respectively. The bands are generated by varying the fraction of infected individuals that can be confirmed as infected by testing, $r_c$, between 75\% and 90\%. The other panels show the variation of $f^{min}_e$ with $f_c, f_T$ and $r_c$.

If we take a closer look at eq.~(\ref{eq:need_can}) and the left-most panel of figure~\ref{fig:f_e_p_t} we see that even with a modest probability of transmission $p_t$ (e.g. about 30\%) quite a large fraction of the population (about 40\% -- 60\%) needs to be enrolled in automated contact tracing even when we assume almost all of them will be actively participating in confirming when they get infected. Assuming all the traced contacts within radius $r_0$ lasting for more than $t_0$ period of time are going to be infected is equivalent to stating $p_t = 100\%$. From the panel on the right, we can see how a fall in the fraction of individuals that confirm that they are infected, $f_c$, can increase $f^{min}_e$. Even with quite low values of $p_t$ nearly half the population needs to be enrolled in automated contact tracing.

Let us try to understand why the effectiveness of automated contact tracing seems to drop so drastically with the enrollment fraction $f_e$. From the left-hand side of eq.~(\ref{eq:need_can}) we see that the effectiveness of automated contact tracing drops as $f_e^2$. We see that $\eta$ drops to 64\% when $f_e = 0.8f^{min}_e$ and 25\% when $f_e = 0.5f^{min}_e$. This non-linearity exists because $f_e$ not only reduces the number of infected individuals who can report their status but also the number of individuals who can receive a notification that they have come in contact with an infected person. The primary reason behind this is the fact that the automated contact tracing depends on voluntary participation whereas manual contact tracing or the use of CCTV, credit card information or identity logging at visited location to trace contact are not voluntary in their current form of implementation\footnote{These effectively makes $f_e$ close to 100\% for both those who have been diagnosed as infected and their contacts.}.

Furthermore, as seen in figure~\ref{fig:f_e_p_t}, when the percentage of infected individuals who report that they have been tested as infected, $f_c$, is lower than 100\%,  automated contact tracing becomes even less effective. In addition, the percentage of cases that can actually be detected, $r_c$, will realistically be less than 100\% for SARS-CoV-2 because of the prevalence of subclinical cases that will escape detection and other clinical factors.

\section*{Assisted contact tracing}

The necessary scale of implementation of automated contact tracing appears to be too large for it to be considered an effective measure to slow down the ongoing pandemic. 
For automated contact tracing to be a viable option, $f^{min}_e$ has to be as low as possible. To achieve this either the product $f_Tp_t$ needs to be decreased or the product $f_cr_c$ needs to be increased as seen from eq.~(\ref{eq:min_enroll}).

\begin{itemize}
\setlength\itemsep{0em}
    \item Both $f_T$ and $p_t$ depend on the dynamics of the disease spreading amongst humans. The fraction of traced cases that need to be quarantined to stop the spread of the disease $f_T$ can be reduced by extensive monitoring of the disease to make sure infected cases are isolated as soon as possible and their contacts are traced. Even a day or two of delays can increase $f_T$ making automated contact tracing ineffective~\cite{Ferrettieabb6936}.
    \item Variations in $p_t$ can be caused by several factors some of which are controllable. Since $p_t$ depends on the contagiousness of the disease and any protective measures taken against the spread of the infection, $p_t$ can be reduced by measures of limited social distancing, the use of PPE and raising public awareness about the contagiousness of COVID-19. This can pose a significant challenge in densely populated regions and regions with poor living conditions and might lead to the breakdown of the applicability of automated contact tracing.
    \item $f_c$ is somewhat more difficult to control assuming the reporting of those who are confirmed as infected is voluntary. This can only be increased by increasing the population's willingness to contribute to automated contact tracing.
    \item $r_c$ is the parameter that is least under control since without very large-scale testing, asymptomatic and mildly symptomatic cases will be difficult to find. This is especially true if the infection can spread by means other than proximity alone as might be the case for SARS-CoV-2~\cite{Santarpia2020.03.23.20039446,doi:10.1056/NEJMc2004973,Guo:2020ww}.
\end{itemize}

Thus we see that a combination of several measures along with a large participation of the population in contact tracing would be the optimal solution for avoiding extensive population-wide social distancing measures and reducing the cost to the economy and well-being of a nation and also allow for greater freedom of movement during a pandemic.

\section*{Discussion}

In our analysis, we have inclined towards an optimistic picture of the spread of SAR-CoV-2. 
We have considered only spreading due to proximity and not considered other means of spreading like contaminated surfaces and aerosol that are common for SARS-CoV-2~\cite{Santarpia2020.03.23.20039446,doi:10.1056/NEJMc2004973,Guo:2020ww} and can increase $p_t$. In figure~\ref{fig:f_e_p_t} we have taken a minimum $r_c$ of 75\% when this can be even lower if widespread testing is not conducted to identify subclinical cases that can go undetected. We have also neglected the requirement for tracing secondary or tertiary contacts. In addition, we have also ignored events where a large number of individuals are infected in very a crowded location like public events for which thresholds like $r_0$ and $t_0$ need to be modified. Despite this optimistic picture, our analysis shows that a majority of the population has to enroll and actively participate in automated contact tracing for the measure to work in the absence of active social distancing measures. 

We have not addressed the sociological aspect of selection bias in the enrollment process. Diversity in socio-economic conditions, awareness of technology and willingness to participate in a community effort will create variation in representations amongst the population. This can lead to the most vulnerable in society getting the least benefit from the implementation of automated contact tracing. Addressing the challenges of implementing automated contact tracing in developing nations where the necessary technologies might not be accessible to a large fraction of the population lies beyond the scope of this work.

We have shown that in real-world scenarios, automated contact tracing alone cannot contain a pandemic driven by a pathogen like the SARS-CoV-2. Advocating it as such can lead to exasperating the spread of the pathogen. The primary reasons why such a strategy will not work as effectively as projected for SARS-CoV-2 is because of a large degree of spreading from presymptomatic and subclinical hosts, and the rapidity with which the virus spreads through proximity alone if no additional measures are taken to mitigate the spread. All of these conjugated with the vulnerability of automated contact tracing to insufficient sampling due to limited participation amongst the population and possibly incomplete reporting of infected cases will lead to reduction in the efficacy of automated contact tracing. A small fraction of the population being infected with SARS-CoV-2 can quickly lead to a majority of the population being needed to participate in the program.

We put together all the factors of concern and show that they follow a simple relationship. We further discussed how factors like the transmission probability $p_t$ should be reduced and the fraction of infected individuals that test positive, $r_c$, should be increased to assist in reducing the burden on automated contact tracing while keeping the entire process voluntary. The strength of our model lies in the fact that we separate the various parameters that individually contribute to the efficacy of automated contact tracing. This allows for individually addressing each parameter through improved clinical intervention, logistics, mitigation strategies and public awareness of automated contact tracing to increase adoption of the method. While our focus in this paper is to address the feasibility of automated contact tracing for containing the spread of SARS-CoV-2, eq.~(\ref{eq:min_enroll}) can be applied for using automated contact tracing to contain other pathogens too. Our analysis is also independent of the methods of implementation of automated contact tracing and the definitions of $r_0$ and $t_0$. Therefore, our approach is quite general.

During the final stages of this work, a similar result was reached by the authors of~\cite{2020arXiv200407237B} using a branching process model and arguments from statistical mechanics. They reached a similar conclusion as we do in our paper showing that nearly 75\% to 95\% of the population need to participate in automated contact tracing for it to be effective. The results in their work corresponds to ours when $p_t=f_c=r_c=1$ or $f_e = \sqrt{f_T}$. A more informed approach based to contact tracing has also been suggested which leads to a lower fraction of the population needing to be enrolled based on a probabilistic model disease spread~\cite{Guttal2020.04.26.20080648}.

The trust in contact tracing stems from the effectiveness with which it was used to contain pathogens like Ebola, SARS-CoV and MERS-CoV. However, the dynamics of the spread of SARS-CoV-2 is very different from these pathogens. Hence, the effectiveness of contact tracing in stopping the spread of these pathogens should not be seen as a validation of the effectiveness of automated contact tracing for SARS-CoV-2. To make automated contact tracing work, a majority of the population has to enroll for this service and actively participate in it. If this cannot be established then other measures of mitigating the spread of SARS-CoV-2 should be implemented in addition. As can be seen by the success of several nations in containing the spread of COVID-19, only a judicious combination of contact tracing with measures such as partial social distancing, wide use of PPE and dissemination of information about the disease can prove to be effective in slowing down the spread of the ongoing pandemic.

\section*{Author Contributions}
Hyunju Kim and Ayan Paul contributed equally to this work.

\section*{Acknowledgements}
We would like to thank Paul Davies, Christophe Grojean, Luca Silvestrini and Melinda Varga for comments and suggestion.

\section*{Data Accessibility}
All data necessary for this work is provided in the paper.

\section*{Funding Statement}
This work was partially supported by a grant from John Templeton Foundation. The opinions/conclusions presented in this publication are those of the author(s) and do not necessarily reflect the views of John Templeton Foundation.

\bibliography{sample}

\begin{thebibliography}{10}
\urlstyle{rm}
\expandafter\ifx\csname url\endcsname\relax
  \def\url#1{\texttt{#1}}\fi
\expandafter\ifx\csname urlprefix\endcsname\relax\def\urlprefix{URL: }\fi
\expandafter\ifx\csname doiprefix\endcsname\relax\def\doiprefix{DOI: }\fi
\providecommand{\bibinfo}[2]{#2}
\providecommand{\eprint}[2][]{\href{https://arxiv.org/abs/#2}{arXiv:#2}}
\providecommand{\url}[2]{\href{https://arxiv.org/abs/#2}{arXiv:#2}}

\bibitem{Kisslereabb5793}
\bibinfo{author}{Kissler, S.~M.}, \bibinfo{author}{Tedijanto, C.},
  \bibinfo{author}{Goldstein, E.}, \bibinfo{author}{Grad, Y.~H.} \&
  \bibinfo{author}{Lipsitch, M.}
\newblock \bibinfo{journal}{\bibinfo{title}{Projecting the transmission
  dynamics of sars-cov-2 through the postpandemic period}}.
\newblock {\emph{\JournalTitle{Science}}} \textbf{\bibinfo{volume}{368}},
  \bibinfo{pages}{860 -- 868},
  \doiprefix\href{https://doi.org/10.1126/science.abb5793}{10.1126/science.abb5793}
  (\bibinfo{year}{2020}).

\bibitem{10.1371/journal.pntd.0006762}
\bibinfo{author}{Swanson, K.~C.} \emph{et~al.}
\newblock \bibinfo{journal}{\bibinfo{title}{Contact tracing performance during
  the ebola epidemic in liberia, 2014-2015}}.
\newblock {\emph{\JournalTitle{PLOS Neglected Tropical Diseases}}}
  \textbf{\bibinfo{volume}{12}}, \bibinfo{pages}{1--14},
  \doiprefix\href{https://doi.org/10.1371/journal.pntd.0006762}{10.1371/journal.pntd.0006762}
  (\bibinfo{year}{2018}).

\bibitem{doi:10.1142/S1793524518500936}
\bibinfo{author}{Berge, T.}, \bibinfo{author}{Ouemba~Tass\'e, A.~J.},
  \bibinfo{author}{Tenkam, H.~M.} \& \bibinfo{author}{Lubuma, J.}
\newblock \bibinfo{journal}{\bibinfo{title}{Mathematical modeling of contact
  tracing as a control strategy of ebola virus disease}}.
\newblock {\emph{\JournalTitle{International Journal of Biomathematics}}}
  \textbf{\bibinfo{volume}{11}}, \bibinfo{pages}{1850093},
  \doiprefix\href{https://doi.org/10.1142/S1793524518500936}{10.1142/S1793524518500936}
  (\bibinfo{year}{2018}).

\bibitem{BROWNE201533}
\bibinfo{author}{Browne, C.}, \bibinfo{author}{Gulbudak, H.} \&
  \bibinfo{author}{Webb, G.}
\newblock \bibinfo{journal}{\bibinfo{title}{Modeling contact tracing in
  outbreaks with application to ebola}}.
\newblock {\emph{\JournalTitle{Journal of Theoretical Biology}}}
  \textbf{\bibinfo{volume}{384}}, \bibinfo{pages}{33 -- 49},
  \doiprefix\href{https://doi.org/10.1016/j.jtbi.2015.08.004}{10.1016/j.jtbi.2015.08.004}
  (\bibinfo{year}{2015}).

\bibitem{1551-0018_2018_5_1165}
\bibinfo{author}{Narges Montazeri~Shahtori, C.~S., Tanvir~Ferdousi} \&
  \bibinfo{author}{Sahneh, F.~D.}
\newblock \bibinfo{journal}{\bibinfo{title}{Quantifying the impact of
  early-stage contact tracing on controlling ebola diffusion}}.
\newblock {\emph{\JournalTitle{Mathematical Biosciences \& Engineering}}}
  \textbf{\bibinfo{volume}{15}}, \bibinfo{pages}{1165},
  \doiprefix\href{https://doi.org/10.3934/mbe.2018053}{10.3934/mbe.2018053}
  (\bibinfo{year}{2018}).

\bibitem{KWOK2019186}
\bibinfo{author}{Kwok, K.~O.} \emph{et~al.}
\newblock \bibinfo{journal}{\bibinfo{title}{Epidemic models of contact tracing:
  Systematic review of transmission studies of severe acute respiratory
  syndrome and middle east respiratory syndrome}}.
\newblock {\emph{\JournalTitle{Computational and Structural Biotechnology
  Journal}}} \textbf{\bibinfo{volume}{17}}, \bibinfo{pages}{186 -- 194},
  \doiprefix\href{https://doi.org/10.1016/j.csbj.2019.01.003}{10.1016/j.csbj.2019.01.003}
  (\bibinfo{year}{2019}).

\bibitem{OKOLIE2020108320}
\bibinfo{author}{Okolie, A.} \& \bibinfo{author}{M\"uller, J.}
\newblock \bibinfo{journal}{\bibinfo{title}{Exact and approximate formulas for
  contact tracing on random trees}}.
\newblock {\emph{\JournalTitle{Mathematical Biosciences}}}
  \textbf{\bibinfo{volume}{321}}, \bibinfo{pages}{108320},
  \doiprefix\href{https://doi.org/10.1016/j.mbs.2020.108320}{10.1016/j.mbs.2020.108320}
  (\bibinfo{year}{2020}).

\bibitem{10.1371/journal.pone.0000012}
\bibinfo{author}{Klinkenberg, D.}, \bibinfo{author}{Fraser, C.} \&
  \bibinfo{author}{Heesterbeek, H.}
\newblock \bibinfo{journal}{\bibinfo{title}{The effectiveness of contact
  tracing in emerging epidemics}}.
\newblock {\emph{\JournalTitle{PLOS ONE}}} \textbf{\bibinfo{volume}{1}},
  \bibinfo{pages}{1--7},
  \doiprefix\href{https://doi.org/10.1371/journal.pone.0000012}{10.1371/journal.pone.0000012}
  (\bibinfo{year}{2006}).

\bibitem{10.1093/cid/civ951}
\bibinfo{author}{Corman, V.~M.} \emph{et~al.}
\newblock \bibinfo{journal}{\bibinfo{title}{{Viral Shedding and Antibody
  Response in 37 Patients With Middle East Respiratory Syndrome Coronavirus
  Infection}}}.
\newblock {\emph{\JournalTitle{Clinical Infectious Diseases}}}
  \textbf{\bibinfo{volume}{62}}, \bibinfo{pages}{477--483},
  \doiprefix\href{https://doi.org/10.1093/cid/civ951}{10.1093/cid/civ951}
  (\bibinfo{year}{2015}).

\bibitem{Chowell:2015tn}
\bibinfo{author}{Chowell, G.} \emph{et~al.}
\newblock \bibinfo{journal}{\bibinfo{title}{{Transmission characteristics of
  MERS and SARS in the healthcare setting: a comparative study}}}.
\newblock {\emph{\JournalTitle{BMC Medicine}}} \textbf{\bibinfo{volume}{13}},
  \bibinfo{pages}{210},
  \doiprefix\href{https://doi.org/10.1186/s12916-015-0450-0}{10.1186/s12916-015-0450-0}
  (\bibinfo{year}{2015}).

\bibitem{Fraser6146}
\bibinfo{author}{Fraser, C.}, \bibinfo{author}{Riley, S.},
  \bibinfo{author}{Anderson, R.~M.} \& \bibinfo{author}{Ferguson, N.~M.}
\newblock \bibinfo{journal}{\bibinfo{title}{Factors that make an infectious
  disease outbreak controllable}}.
\newblock {\emph{\JournalTitle{Proceedings of the National Academy of
  Sciences}}} \textbf{\bibinfo{volume}{101}}, \bibinfo{pages}{6146--6151},
  \doiprefix\href{https://doi.org/10.1073/pnas.0307506101}{10.1073/pnas.0307506101}
  (\bibinfo{year}{2004}).

\bibitem{Omrani:2013fp}
\bibinfo{author}{Omrani, A.~S.} \emph{et~al.}
\newblock \bibinfo{journal}{\bibinfo{title}{{A family cluster of Middle East
  Respiratory Syndrome Coronavirus infections related to a likely unrecognized
  asymptomatic or mild case}}}.
\newblock {\emph{\JournalTitle{International Journal of Infectious Diseases}}}
  \textbf{\bibinfo{volume}{17}}, \bibinfo{pages}{e668--e672},
  \doiprefix\href{https://doi.org/10.1016/j.ijid.2013.07.001}{10.1016/j.ijid.2013.07.001}
  (\bibinfo{year}{2013}).

\bibitem{REWAR2014444}
\bibinfo{author}{Rewar, S.} \& \bibinfo{author}{Mirdha, D.}
\newblock \bibinfo{journal}{\bibinfo{title}{Transmission of ebola virus
  disease: An overview}}.
\newblock {\emph{\JournalTitle{Annals of Global Health}}}
  \textbf{\bibinfo{volume}{80}}, \bibinfo{pages}{444 -- 451},
  \doiprefix\href{https://doi.org/10.1016/j.aogh.2015.02.005}{10.1016/j.aogh.2015.02.005}
  (\bibinfo{year}{2014}).

\bibitem{Lau:2010tv}
\bibinfo{author}{Lau, L. L.~H.} \emph{et~al.}
\newblock \bibinfo{journal}{\bibinfo{title}{{Viral shedding and clinical
  illness in naturally acquired influenza virus infections}}}.
\newblock {\emph{\JournalTitle{The Journal of infectious diseases}}}
  \textbf{\bibinfo{volume}{201}}, \bibinfo{pages}{1509--1516},
  \doiprefix\href{https://doi.org/10.1086/652241}{10.1086/652241}
  (\bibinfo{year}{2010}).

\bibitem{Santarpia2020.03.23.20039446}
\bibinfo{author}{Santarpia, J.~L.} \emph{et~al.}
\newblock \bibinfo{journal}{\bibinfo{title}{{Aerosol and surface contamination
  of SARS-CoV-2 observed in quarantine and isolation care}}}.
\newblock {\emph{\JournalTitle{Scientific Reports}}}
  \textbf{\bibinfo{volume}{10}}, \bibinfo{pages}{12732},
  \doiprefix\href{https://doi.org/10.1038/s41598-020-69286-3}{10.1038/s41598-020-69286-3}
  (\bibinfo{year}{2020}).

\bibitem{Wang:2020jb}
\bibinfo{author}{Wang, C.}, \bibinfo{author}{Horby, P.~W.},
  \bibinfo{author}{Hayden, F.~G.} \& \bibinfo{author}{Gao, G.~F.}
\newblock \bibinfo{journal}{\bibinfo{title}{{A novel coronavirus outbreak of
  global health concern}}}.
\newblock {\emph{\JournalTitle{The Lancet}}} \textbf{\bibinfo{volume}{395}},
  \bibinfo{pages}{470--473},
  \doiprefix\href{https://doi.org/10.1016/S0140-6736(20)30185-9}{10.1016/S0140-6736(20)30185-9}
  (\bibinfo{year}{2020}).

\bibitem{He:2020ty}
\bibinfo{author}{He, X.} \emph{et~al.}
\newblock \bibinfo{journal}{\bibinfo{title}{{Temporal dynamics in viral
  shedding and transmissibility of COVID-19}}}.
\newblock {\emph{\JournalTitle{Nature Medicine}}}
  \doiprefix\href{https://doi.org/10.1038/s41591-020-0869-5}{10.1038/s41591-020-0869-5}
  (\bibinfo{year}{2020}).

\bibitem{doi:10.1056/NEJMc2004973}
\bibinfo{author}{van Doremalen, N.} \emph{et~al.}
\newblock \bibinfo{journal}{\bibinfo{title}{Aerosol and surface stability of
  sars-cov-2 as compared with sars-cov-1}}.
\newblock {\emph{\JournalTitle{New England Journal of Medicine}}}
  \textbf{\bibinfo{volume}{382}}, \bibinfo{pages}{1564--1567},
  \doiprefix\href{https://doi.org/10.1056/NEJMc2004973}{10.1056/NEJMc2004973}
  (\bibinfo{year}{2020}).

\bibitem{Guo:2020ww}
\bibinfo{author}{Guo, Z.-D.} \emph{et~al.}
\newblock \bibinfo{journal}{\bibinfo{title}{{Aerosol and Surface Distribution
  of Severe Acute Respiratory Syndrome Coronavirus 2 in Hospital Wards, Wuhan,
  China, 2020}}}.
\newblock {\emph{\JournalTitle{Emerging Infectious Disease journal}}}
  \textbf{\bibinfo{volume}{26}},
  \doiprefix\href{https://doi.org/10.3201/eid2607.200885}{10.3201/eid2607.200885}
  (\bibinfo{year}{2020}).

\bibitem{Zhou:2020wk}
\bibinfo{author}{Zhou, P.} \emph{et~al.}
\newblock \bibinfo{journal}{\bibinfo{title}{{A pneumonia outbreak associated
  with a new coronavirus of probable bat origin}}}.
\newblock {\emph{\JournalTitle{Nature}}} \textbf{\bibinfo{volume}{579}},
  \bibinfo{pages}{270--273},
  \doiprefix\href{https://doi.org/10.1038/s41586-020-2012-7}{10.1038/s41586-020-2012-7}
  (\bibinfo{year}{2020}).

\bibitem{Wrapp1260}
\bibinfo{author}{Wrapp, D.} \emph{et~al.}
\newblock \bibinfo{journal}{\bibinfo{title}{Cryo-em structure of the 2019-ncov
  spike in the prefusion conformation}}.
\newblock {\emph{\JournalTitle{Science}}} \textbf{\bibinfo{volume}{367}},
  \bibinfo{pages}{1260--1263},
  \doiprefix\href{https://doi.org/10.1126/science.abb2507}{10.1126/science.abb2507}
  (\bibinfo{year}{2020}).

\bibitem{CHEN2020135}
\bibinfo{author}{Chen, Y.}, \bibinfo{author}{Guo, Y.}, \bibinfo{author}{Pan,
  Y.} \& \bibinfo{author}{Zhao, Z.~J.}
\newblock \bibinfo{journal}{\bibinfo{title}{Structure analysis of the receptor
  binding of 2019-ncov}}.
\newblock {\emph{\JournalTitle{Biochemical and Biophysical Research
  Communications}}} \textbf{\bibinfo{volume}{525}}, \bibinfo{pages}{135 --
  140},
  \doiprefix\href{https://doi.org/10.1016/j.bbrc.2020.02.071}{10.1016/j.bbrc.2020.02.071}
  (\bibinfo{year}{2020}).

\bibitem{Lieabb3221}
\bibinfo{author}{Li, R.} \emph{et~al.}
\newblock \bibinfo{journal}{\bibinfo{title}{Substantial undocumented infection
  facilitates the rapid dissemination of novel coronavirus (sars-cov2)}}.
\newblock {\emph{\JournalTitle{Science}}}
  \doiprefix\href{https://doi.org/10.1126/science.abb3221}{10.1126/science.abb3221}
  (\bibinfo{year}{2020}).

\bibitem{doi:10.1056/NEJMoa2001316}
\bibinfo{author}{Li, Q.} \emph{et~al.}
\newblock \bibinfo{journal}{\bibinfo{title}{Early transmission dynamics in
  wuhan, china, of novel coronavirus-infected pneumonia}}.
\newblock {\emph{\JournalTitle{New England Journal of Medicine}}}
  \textbf{\bibinfo{volume}{382}}, \bibinfo{pages}{1199--1207},
  \doiprefix\href{https://doi.org/10.1056/NEJMoa2001316}{10.1056/NEJMoa2001316}
  (\bibinfo{year}{2020}).

\bibitem{Wu:2020ix}
\bibinfo{author}{Wu, J.~T.}, \bibinfo{author}{Leung, K.} \&
  \bibinfo{author}{Leung, G.~M.}
\newblock \bibinfo{journal}{\bibinfo{title}{{Nowcasting and forecasting the
  potential domestic and international spread of the 2019-nCoV outbreak
  originating in Wuhan, China: a modelling study}}}.
\newblock {\emph{\JournalTitle{The Lancet}}} \textbf{\bibinfo{volume}{395}},
  \bibinfo{pages}{689--697},
  \doiprefix\href{https://doi.org/10.1016/S0140-6736(20)30260-9}{10.1016/S0140-6736(20)30260-9}
  (\bibinfo{year}{2020}).

\bibitem{10.2807/1560-7917.ES.2020.25.4.2000058}
\bibinfo{author}{Riou, J.} \& \bibinfo{author}{Althaus, C.~L.}
\newblock \bibinfo{journal}{\bibinfo{title}{Pattern of early human-to-human
  transmission of wuhan 2019 novel coronavirus (2019-ncov), december 2019 to
  january 2020}}.
\newblock {\emph{\JournalTitle{Eurosurveillance}}}
  \textbf{\bibinfo{volume}{25}},
  \doiprefix\href{https://doi.org/10.2807/1560-7917.ES.2020.25.4.2000058}{10.2807/1560-7917.ES.2020.25.4.2000058}
  (\bibinfo{year}{2020}).

\bibitem{Du:2020va}
\bibinfo{author}{Du, Z.} \emph{et~al.}
\newblock \bibinfo{journal}{\bibinfo{title}{{Risk for Transportation of 2019
  Novel Coronavirus Disease from Wuhan to Other Cities in China}}}.
\newblock {\emph{\JournalTitle{Emerging Infectious Disease journal}}}
  \textbf{\bibinfo{volume}{26}},
  \doiprefix\href{https://doi.org/10.3201/eid2605.200146}{10.3201/eid2605.200146}
  (\bibinfo{year}{2020}).

\bibitem{Sanche:2020wm}
\bibinfo{author}{Sanche, S.} \emph{et~al.}
\newblock \bibinfo{journal}{\bibinfo{title}{{High Contagiousness and Rapid
  Spread of Severe Acute Respiratory Syndrome Coronavirus 2}}}.
\newblock {\emph{\JournalTitle{Emerging Infectious Disease journal}}}
  \textbf{\bibinfo{volume}{26}},
  \doiprefix\href{https://doi.org/10.3201/eid2607.200282}{10.3201/eid2607.200282}
  (\bibinfo{year}{2020}).

\bibitem{ALRAEEI2020}
\bibinfo{author}{Al-Raeei, M.}
\newblock \bibinfo{journal}{\bibinfo{title}{{The basic reproduction number of
  the new coronavirus pandemic with mortality for India, the Syrian Arab
  Republic, the United States, Yemen, China, France, Nigeria and Russia with
  different rate of cases}}}.
\newblock {\emph{\JournalTitle{Clinical Epidemiology and Global Health}}}
  \doiprefix\href{https://doi.org/10.1016/j.cegh.2020.08.005}{10.1016/j.cegh.2020.08.005}
  (\bibinfo{year}{2020}).

\bibitem{LloydSmith:2005ue}
\bibinfo{author}{Lloyd-Smith, J.~O.}, \bibinfo{author}{Schreiber, S.~J.},
  \bibinfo{author}{Kopp, P.~E.} \& \bibinfo{author}{Getz, W.~M.}
\newblock \bibinfo{journal}{\bibinfo{title}{{Superspreading and the effect of
  individual variation on disease emergence}}}.
\newblock {\emph{\JournalTitle{Nature}}} \textbf{\bibinfo{volume}{438}},
  \bibinfo{pages}{355--359},
  \doiprefix\href{https://doi.org/10.1038/nature04153}{10.1038/nature04153}
  (\bibinfo{year}{2005}).

\bibitem{Hellewell:2020ek}
\bibinfo{author}{Hellewell, J.} \emph{et~al.}
\newblock \bibinfo{journal}{\bibinfo{title}{{Feasibility of controlling
  COVID-19 outbreaks by isolation of cases and contacts}}}.
\newblock {\emph{\JournalTitle{The Lancet Global Health}}}
  \textbf{\bibinfo{volume}{8}}, \bibinfo{pages}{e488--e496},
  \doiprefix\href{https://doi.org/10.1016/S2214-109X(20)30074-7}{10.1016/S2214-109X(20)30074-7}
  (\bibinfo{year}{2020}).

\bibitem{10.12688/wellcomeopenres.15842.1}
\bibinfo{author}{Endo, A.}, \bibinfo{author}{Abbott, S.},
  \bibinfo{author}{Kucharski, A.} \& \bibinfo{author}{Funk, S.}
\newblock \bibinfo{journal}{\bibinfo{title}{Estimating the overdispersion in
  covid-19 transmission using outbreak sizes outside china [version 1; peer
  review: awaiting peer review]}}.
\newblock {\emph{\JournalTitle{Wellcome Open Research}}}
  \textbf{\bibinfo{volume}{5}},
  \doiprefix\href{https://doi.org/10.12688/wellcomeopenres.15842.1}{10.12688/wellcomeopenres.15842.1}
  (\bibinfo{year}{2020}).

\bibitem{Keeling2020.02.14.20023036}
\bibinfo{author}{Keeling, M.~J.}, \bibinfo{author}{Hollingsworth, T.~D.} \&
  \bibinfo{author}{Read, J.~M.}
\newblock \bibinfo{journal}{\bibinfo{title}{Efficacy of contact tracing for the
  containment of the 2019 novel coronavirus (covid-19)}}.
\newblock {\emph{\JournalTitle{Journal of Epidemiology \& Community Health}}}
  \textbf{\bibinfo{volume}{74}}, \bibinfo{pages}{861--866},
  \doiprefix\href{https://doi.org/10.1136/jech-2020-214051}{10.1136/jech-2020-214051}
  (\bibinfo{year}{2020}).

\bibitem{Ferrettieabb6936}
\bibinfo{author}{Ferretti, L.} \emph{et~al.}
\newblock \bibinfo{journal}{\bibinfo{title}{Quantifying sars-cov-2 transmission
  suggests epidemic control with digital contact tracing}}.
\newblock {\emph{\JournalTitle{Science}}}
  \doiprefix\href{https://doi.org/10.1126/science.abb6936}{10.1126/science.abb6936}
  (\bibinfo{year}{2020}).

\bibitem{Pung:2020ef}
\bibinfo{author}{Pung, R.} \emph{et~al.}
\newblock \bibinfo{journal}{\bibinfo{title}{{Investigation of three clusters of
  COVID-19 in Singapore: implications for surveillance and response
  measures}}}.
\newblock {\emph{\JournalTitle{The Lancet}}} \textbf{\bibinfo{volume}{395}},
  \bibinfo{pages}{1039--1046},
  \doiprefix\href{https://doi.org/10.1016/S0140-6736(20)30528-6}{10.1016/S0140-6736(20)30528-6}
  (\bibinfo{year}{2020}).

\bibitem{info:doi/10.2196/19540}
\bibinfo{author}{Chen, C.-M.} \emph{et~al.}
\newblock \bibinfo{journal}{\bibinfo{title}{Containing covid-19 among 627,386
  persons in contact with the diamond princess cruise ship passengers who
  disembarked in taiwan: Big data analytics}}.
\newblock {\emph{\JournalTitle{J Med Internet Res}}}
  \textbf{\bibinfo{volume}{22}}, \bibinfo{pages}{e19540},
  \doiprefix\href{https://doi.org/10.2196/19540}{10.2196/19540}
  (\bibinfo{year}{2020}).

\bibitem{10.1001/jama.2020.6602}
\bibinfo{author}{Park, S.}, \bibinfo{author}{Choi, G.~J.} \&
  \bibinfo{author}{Ko, H.}
\newblock \bibinfo{journal}{\bibinfo{title}{{Information Technology-Based
  Tracing Strategy in Response to COVID-19 in South Korea-Privacy
  Controversies}}}.
\newblock {\emph{\JournalTitle{JAMA}}} \textbf{\bibinfo{volume}{323}},
  \bibinfo{pages}{2129--2130},
  \doiprefix\href{https://doi.org/10.1001/jama.2020.6602}{10.1001/jama.2020.6602}
  (\bibinfo{year}{2020}).

\bibitem{Daym1375}
\bibinfo{author}{Day, M.}
\newblock \bibinfo{journal}{\bibinfo{title}{Covid-19: four fifths of cases are
  asymptomatic, china figures indicate}}.
\newblock {\emph{\JournalTitle{BMJ}}} \textbf{\bibinfo{volume}{369}},
  \doiprefix\href{https://doi.org/10.1136/bmj.m1375}{10.1136/bmj.m1375}
  (\bibinfo{year}{2020}).

\bibitem{Verity:2020cu}
\bibinfo{author}{Verity, R.} \emph{et~al.}
\newblock \bibinfo{journal}{\bibinfo{title}{{Estimates of the severity of
  coronavirus disease 2019: a model-based analysis}}}.
\newblock {\emph{\JournalTitle{The Lancet Infectious Diseases}}}
  \doiprefix\href{https://doi.org/10.1016/S1473-3099(20)30243-7}{10.1016/S1473-3099(20)30243-7}
  (\bibinfo{year}{2020}).

\bibitem{MORAWSKA2020105832}
\bibinfo{author}{Morawska, L.} \emph{et~al.}
\newblock \bibinfo{journal}{\bibinfo{title}{{How can airborne transmission of
  COVID-19 indoors be minimised?}}}
\newblock {\emph{\JournalTitle{Environment International}}}
  \textbf{\bibinfo{volume}{142}}, \bibinfo{pages}{105832},
  \doiprefix\href{https://doi.org/10.1016/j.envint.2020.105832}{10.1016/j.envint.2020.105832}
  (\bibinfo{year}{2020}).

\bibitem{Ferretti2020.09.04.20188516}
\bibinfo{author}{Ferretti, L.} \emph{et~al.}
\newblock \bibinfo{journal}{\bibinfo{title}{{The timing of COVID-19
  transmission}}}.
\newblock {\emph{\JournalTitle{medRxiv}}}
  \doiprefix\href{https://doi.org/10.1101/2020.09.04.20188516}{10.1101/2020.09.04.20188516}
  (\bibinfo{year}{2020}).

\bibitem{2020arXiv200407237B}
\bibinfo{author}{{Bulchandani}, V.~B.}, \bibinfo{author}{{Shivam}, S.},
  \bibinfo{author}{{Moudgalya}, S.} \& \bibinfo{author}{{Sondhi}, S.~L.}
\newblock \bibinfo{journal}{\bibinfo{title}{{Digital Herd Immunity and
  COVID-19}}}.
\newblock {\emph{\JournalTitle{arXiv e-prints}}}  (\bibinfo{year}{2020}).
\newblock \eprint{2004.07237}.

\bibitem{Guttal2020.04.26.20080648}
\bibinfo{author}{Guttal, V.}, \bibinfo{author}{Krishna, S.} \&
  \bibinfo{author}{Siddharthan, R.}
\newblock \bibinfo{journal}{\bibinfo{title}{{Risk assessment via layered mobile
  contact tracing for epidemiological intervention}}}.
\newblock {\emph{\JournalTitle{medRxiv}}}
  \doiprefix\href{https://doi.org/10.1101/2020.04.26.20080648}{10.1101/2020.04.26.20080648}
  (\bibinfo{year}{2020}).

\end{thebibliography}




\end{document}